\documentclass{JINST}

\usepackage{amsmath,amsfonts,amssymb,fontenc,times,mathptmx,graphicx,graphics}

\title{Asymmetries in Silicon Microstrip Response Function and Lorentz Angle }
\author{Gregorio Landi$^a$\thanks{Corresponding
author.}~, and  Giovanni E. Landi$^b$\\
\\
\llap{$^a$} Dipartimento di Fisica e Astronomia, Universita' di Firenze,\\
Largo E. Fermi 2 50125 Firenze Italy\\
and INFN, Sezione di Firenze, \\
Firenze,Italy\\
E-mail: \email{landi@fi.infn.it}\\
\\
\llap{$^b$} UBICA s.r.l.,\\
Via S. Siro 6/1,\\
Genova, Italy.\\}
\abstract{
An experimental set up, dedicated to isolate an error present in
the $\eta$-algorithm, gave an unexpected result. The average of
a center of gravity algorithm at orthogonal particle incidence turns out
to be non zero.
This non zero average signals an asymmetry in
the response function of the strips, and
introduces a further parameter in the corrections: the shift of
the strip response center of gravity respect its geometrical position.
A strategy to extract this parameter from a standard data
set is discussed. Some simulations with various asymmetric response functions
are explored for this test. The method is able to detect easily the asymmetry
parameters introduced in the simulations. Its robustness is tested
against angular rotations, and we see an almost linear variation
with the angle. This simple property is used to simulate a
determination of a Lorentz angle with and without the
asymmetry of the response function.
}


\begin{document}

\section{Introduction}

In many high-energy physics experiments, arrays of silicon microstrip
detectors are fundamental tools to track charged particles. The excellent position
resolution of these detectors is essential in the event reconstruction.
To obtain the best performance, the role played by the position-reconstruction
algorithms becomes crucial. For example, the final alignments
are corrected with track reconstructions;  any inaccuracy in the
position reconstruction algorithms is systematically diffused
to all the data. The use of reconstruction algorithms in the detector
alignment and in the data creates correlations that renders almost
impossible to verify their consistency.
Thus, an {\em a priori} exploration of their systematic errors is
essential.

In a previous article~\cite{landi03} we applied to silicon microstrip
detectors the general equations we developed
in~\cite{landi01,landi02} for the center of gravity (COG)
algorithm. Among the many properties
demonstrated for the COG, we  underlined
the presence of a systematic error in the so called
$\eta$-algorithm~\cite{belau}, when used outside
the symmetry conditions.
The authors in~\cite{belau} recommended the limitation to a symmetric configuration
without demonstration. Thus,
in the last years, the recommendation has been neglected,
and the $\eta$-algorithm has been
used well outside its range of validity. It is easy to
guess the production of many incorrect position reconstructions.

The $\eta$-algorithm improves the COG-algorithm with a global analysis
of a set of equivalent data. Our procedure to define the
$\eta-$algorithm is substantially different from that used in~\cite{belau}.
We deduce it from the
solution of a first order differential equation that has an easy
solution for a uniform distribution of impact points. But,
any first order differential equation always requires an
initial constant, in this case an exact impact point corresponding
to a COG value. This type of datum is never
available excluding some special cases. The initial
constant is easily selected for symmetrical configurations, and is
zero with the definitions of~\cite{belau}. For unsymmetrical
configuration, for example at
non-orthogonal incidence angles, an angle dependent shift is
produced by the use of the zero constant of the
symmetric case. The shift depends on the form of the signal
distribution. Thus, detectors aligned with minimum
ionizing particles (MIP) could show non alignments with heavy
ions (in reality there are non alignments in both cases).
Similar apparent shift of a detector could be induced by the
modification of the depleting tension or any other deformation of
the signal distribution. Simulations show shifts
greater than the root mean square (RMS) error in some directions,
and always larger than the full width half maximum (FWHM)  of the error
distributions. In ref.~\cite{landi03}, we
demonstrate a method to correct it.

We have to underline the importance of the $\eta$-algorithm in improving the
position reconstructions. The comparison of the RMS-error of the COG
and $\eta$-algorithm does not show dramatic differences in favor of the latter,
as the comparison of the FWHM. The reason of the small sensitivity of the RMS-error
to the improvement of the $\eta$-algorithm is connected to the non-linear dependence of the
two algorithms from their component stochastic variables. As it is well known,
non-linearities introduce drastic deviation from the gaussian distributions toward
slow decreasing probability distributions. The Cauchy distribution is a
typical member of this class. These non-gaussian distributions tend to have
infinite variances as the Cauchy distribution.
In this case, the RMS-error is essentially limited the selection strategy of the finite sample
and it is insensible to the quality of the reconstruction algorithms. On the contrary
the FWHM saves its sensitivity.

In a test beam with a set of sensors  of the PAMELA
tracker~\cite{PAMELA}, a special set up was exposed to the beam with the aim to
measure the systematic error of the $\eta$-algorithm.
The analysis of the collected data~\cite{vannu} clearly confirms the
presence of an angle-dependent shift, and the correction proposed in ref.~\cite{landi03} is
able to cancel the shift at any measured
angle.

In this work we concentrate the attention on an anomaly
observed on the data of ref.~\cite{vannu} where the average of the COG distribution is
appreciably different from zero for orthogonal particle incidence.
In the absence of magnetic field, the maximal
symmetry is expected for this configuration with the COG probability distribution
symmetric respect to the origin and zero average. The
non zero average could be originated by an asymmetry in the charge drift to
the collection pads or some other (linear) distortion in the
read-out chain.

Our correction to the $\eta$-algorithm works identically for asymmetrical
response functions, but, a further detector parameter must be known:
the COG position of the strip response function. In fact, the COG algorithm assumes that
the strip signals are concentrated in the COG position of the
strip response function. The asymmetry moves the COG
response function from the strip axis, and
this shift must be accounted for in any reconstruction at any angle, not only
in the $\eta$-algorithm.

We have no control on the physics of the showering particle, but, we
suppose to know all the detector parameters, being the detector
production under our control. In practice the situation is not so
simple. Various types of material depositions are performed in
specialized places and slight asymmetries could be easily introduced
during these operations, no visual or electronic inspection can
isolate these defects. In addition to this, subtle asymmetries
could be introduced in the path of the data from the detector
to final user.

Direct measurements could be performed, but they require
auxiliary detectors with resolutions much better than the
tested detectors. It is evident the complexity of this task.
We will tray to estimate the asymmetry from the charge
collected by the strips for
MIP at orthogonal incidence angle. In this way, a good angular measurement
can replace a high resolution position measurements.

In section 2  we give a direct demonstration of the $\eta$-algorithm
correction in general cases to isolate the effects of the asymmetry.
Section 3 is devoted to define our strategy
to estimate the asymmetry parameter of the response functions and to
test it on simulated data with two different type of asymmetry.
Our simulations are tuned on the double sided
silicon microstrip detectors of the type introduced by
ref.~\cite{aleph,L3}, and used in the PAMELA detector. In one side a
strip each two is left unconnected, and it distributes the charge
in a peculiar mode. We call this side floating strip side.
The other side is normal (in the sense that it has no floating
strips).

Section 4 deals with the non orthogonal particle incidence and its
relation with the asymmetry. The angular rotation introduces a simple and almost linear effect
that allows a better determination of the asymmetry. It gives even an indication of
the angular precision to obtain significative results. This sensitivity to the angular
rotation suggests a method to measure the Lorentz angle when a  magnetic field is present.
Here, the effect of a magnetic field an a silicon microstrip detector is simulated as an
effective rotation of the incoming particle direction. A proper  angular rotation is
able to restore the maximal symmetry to the signal distribution. Our method easily
find this condition even in presence of an asymmetry of the response function.
The simulations of this case show an excellent sensitivity of the method.

We are aware that these developments are very formal and complex, but the asymmetry correction
and the Lorentz angle are deeply buried in the properties of the COG algorithm.
It is interesting that analytical developments are able to isolate them and reach the consistency
displayed by the simulations.

\section{Correction of the systematic errors} \label{sec:second}
\subsection{COG averages}

In ref.~\cite{landi01,landi02,landi03} we extensively utilized the
Fourier Transform (FT) and Poisson identity~\cite{libroFT,libroFT2} (or the Shannon sampling theorem).
Now, we will proceed
in a different way that avoids some technical complications
and underlines its generality.

Let us derive the COG average. With the notation
of~\cite{landi01,landi02} and considering all the strips with
a non zero energy, we have the following definition for
the COG ($\tau$ is the strip dimension):
\begin{equation}\label{eq:equation01}
x_g(\varepsilon)=\frac{\sum_{n\in\, \mathbb{Z}}n\tau\
\mathrm{f}(n\tau-\varepsilon)}{\sum_{n\in\, \mathbb{Z}}
\mathrm{f}(n\tau-\varepsilon)}
\end{equation}
where f$(n\tau-\varepsilon)$ is the energy collected by a strip
centered in $n\tau$ for a signal distribution with its COG in
$\varepsilon$ (for any $\varepsilon\in \mathbb{R}$). We use an
infinite sum, but the function f$(n\tau-\varepsilon)$ is expected to go
to zero for a fixed range of its argument (finite support function).
An identical transformation on equation~\ref{eq:equation01} gives:
\begin{equation}\label{eq:equation02}
x_g(\varepsilon)-\varepsilon=\frac{\sum_{n\in\,
\mathbb{Z}}(n\tau-\varepsilon)
\mathrm{f}(n\tau-\varepsilon)}{\sum_{n\in\, \mathbb{Z}}
\mathrm{f}(n\tau-\varepsilon)}\,.
\end{equation}
Equation~\ref{eq:equation02} explicitly shows  the $\tau$-periodicity of
$x_g(\varepsilon)-\varepsilon$ and justifies the use of Fourier
Series (FS). The assumption of absence of signal loss gives a flat
efficiency surface and f$(n\tau-\varepsilon)$ has the sum
rule:
\begin{equation}\label{eq:equation03}
\sum_{n\in\, \mathbb{Z}} \mathrm{f}(n\tau-\varepsilon)=1\ \ \
\forall\varepsilon\in \mathbb{R},
\end{equation}
allowing the suppression of the denominator in equation~\ref{eq:equation02}.
The energy $\mathrm{f}(n\tau-\varepsilon)$ is defined as the
convolution of the strip response function g$(x)$ with the signal
distribution $\varphi(x-\varepsilon)$. The response function g$(x)$ is
centered on the fiducial strip position and $\varphi(x)$ has its COG in
$\varepsilon$:
\begin{equation}\label{eq:equation04}
\mathrm{f}(n\tau-\varepsilon)=\int_{-\infty}^{+\infty}\mathrm{g}(n\tau-x')\varphi(x'-\varepsilon)\mathrm{d}
x'
\end{equation}
With equation~\ref{eq:equation03}, the $\varepsilon$-average on a period $\tau$ of equation~\ref{eq:equation02} acquires an easy aspect. The introduction of the integration variables
$\xi_n=n\tau-\varepsilon$ gives:
\begin{equation*}
\frac{1}{\tau}\int_{-\tau/2}^{+\tau/2}(x_g(\varepsilon)-\varepsilon)\mathrm{d}\varepsilon=\frac{1}{\tau}\sum_{n\in\,
\mathbb{Z}}\int_{n\tau-\tau/2}^{n\tau+\tau/2}\xi_n\
\mathrm{f}(\xi_n)\mathrm{d}\xi_n\,,
\end{equation*}
the sum on $n$ can be absorbed in the definition of the integration limits:
\begin{equation}
\label{eq:equation05}
\frac{1}{\tau}\int_{-\tau/2}^{+\tau/2}(x_g(\varepsilon)-\varepsilon)\mathrm{d}\varepsilon=\frac{1}{\tau}\int_{-\infty}^{+\infty}\xi\
\mathrm{f}(\xi)\mathrm{d}\xi
\end{equation}
Equation~\ref{eq:equation05} is the first momentum of
$\mathrm{f}(x)$, and the convolution theorem for the first
momenta~\cite{libroFT} gives:
\begin{equation*}
\int_{-\infty}^{+\infty}\xi\
\mathrm{f}(\xi)\mathrm{d}\xi=\delta_g\tau+\delta_\varphi
\end{equation*}
Where $\delta_g$ and $\delta_\varphi$ are defined as:
\begin{equation*}
\delta_g=\frac{1}{\tau}\int_{-\infty}^{+\infty}\xi\,\mathrm{g}(\xi)\,\mathrm{d}\xi \ \ \ \ \ \ \ \
\delta_\varphi=\int_{-\infty}^{+\infty}\xi\,\varphi(\xi)\,\mathrm{d}\xi
\end{equation*}
For their normalizations
($\int_{-\infty}^{+\infty}\varphi(\xi)\mathrm{d}\xi=1$ and
$\int_{-\infty}^{+\infty}\mathrm{g}(\xi)\mathrm{d}\xi=\tau$), $\delta_g$
is the COG position of the response function and
$\delta_\varphi$ is the COG position of the signal distribution.
The COG $\delta_\varphi$ is zero for our definition of
$\varepsilon$, and the average of equation~\ref{eq:equation02} remains:
\begin{equation}
\label{eq:equation05_b}
\frac{1}{\tau}\int_{-\tau/2}^{+\tau/2}(x_g(\varepsilon)-\varepsilon)\mathrm{d}\varepsilon=\delta_g\,.
\end{equation}
Equation~\ref{eq:equation05_b} shows that the COG algorithm is a biased
estimator of the impact point. To eliminate this bias,
equation~\ref{eq:equation05_b} imposes that the fiducial strip position
must be coincident with the COG of its response function g$(x)$, in this
case $\delta_g=0$. Any deviation from this condition introduces a
constant shift in the reconstructed position.

In principle, the
extraction of $\delta_g$ from the data is easy, one has to take a
set of (uniform) events, where the values of $\{\varepsilon_j\}$
are known, and to average the differences
$x_g(\varepsilon_j)-\varepsilon_j$. In practice, the value of
$\varepsilon_j$ is very difficult (or impossible) to measure with the due
precision. Thus,
we have to find another strategy to obtain a reasonable
estimation of $\delta_g$ from the data of a standard test beam experiment.

%
%
%
%
Equation~\ref{eq:equation05_b} is evidently valid for a noiseless case.
The data are
surely noisy. Assuming a symmetric additive noise, it is easy to figure out
how it will modify the COG. At fixed  impact point the noise will spread the
data around the noiseless COG value. The symmetry of the noise distribution
induces a symmetric distribution of COG values around the noiseless one and
the averages of the noisy data will converge to the noiseless ones. So, for
a large data sample, our noiseless equations will work identically even
in presence of noise.
\subsection{ The $\eta$-Algorithms}
Let us see how $\delta_g$ modifies the correction of the $\eta$
algorithms. As we proved in refs.~\cite{landi01,landi03},
$\eta$-algorithms may be extended beyond the two strip case used in
ref.~\cite{belau}, and identified as a general property of any COG algorithms.
Due to their strict similarity, we will continue
to call $\eta$-algorithms all these extensions.

The COG algorithms with different numbers of signal strips have
very different properties and systematic errors, and a great care must be
devoted to avoid to mix them. For example, the cuts on small or negative values
of signal strips may produce the mixing. In ref.~\cite{landi03},
2-strips, 3-strips and 4-strips algorithms exhausted our needs,
there we limited to consider incidence angles up to $20^\circ$.
Above $20^\circ$, 5 or more strips are relevant, and other strategies
can be used to reduce these cases to the present developments.

In the simulations, the set of events has $\varepsilon$-values
with a uniform distribution on a strip. This assumption supports
our averages over $\varepsilon$.  As in ref.~\cite{landi03} we
calculate the COG in a reference system bound to the event, we
choose the maximum signal strip.
The experimental events are spread over a large number of
strips. To be consistent with our simulations, we will assume that the set of events
$\{\varepsilon(j)\}$ produces the uniform distribution of points
$\{\sum_{K\in\mathbb{Z}}\varepsilon(j)+K\tau\}$. Thus, on a given strip,
one has the uniform distribution of points
$\{\varepsilon(j)+K_j\tau\}$, where $K_j\tau$ is the distance of
the strip with the impact point $\varepsilon(j)$ from the given
strip. This will be the definition of uniformity of events on a strip.

Let us recall some aspects of the $\eta$ algorithm~\cite{landi03}
to define the notation. Assuming the
existence of a single valued function $x_{gk}(\varepsilon)$
which is randomly sampled by our COG algorithm with $k-$strips
(in the following the index $k$ will indicate the number of strips used
in the algorithm),
the probability to have $x_{gk}$ is:

\begin{equation*}
P(\varepsilon)\,\Big|\frac{\mathrm{d}\varepsilon}{\mathrm{d}
x_{gk}}\Big|=\Gamma(x_{gk})\,,
\end{equation*}
where $P(\varepsilon)$ is the probability to have a value
$\varepsilon$ and $\Gamma(x_{gk})$ is the corresponding
probability for $x_{gk}$. The positivity of the
derivative is reported in ref.~\cite{landi01} and it turns
out that any incoming signal, with average positive signal distribution, has positive derivative.
Assuming a constant probability $P(\varepsilon)=1/\tau$, one arrives to the
first order differential equation:
\begin{equation}\label{eq:equation08}
\frac{1}{\tau}\frac{\mathrm{d}\varepsilon}{\mathrm{d} x_{gk}}=\Gamma(x_{gk})\,.
\end{equation}
The integration of equation~\ref{eq:equation08}
requires an initial constant (i.e., an exact value of the impact
point $\varepsilon(x_{gk})$). For symmetric signal distribution
and symmetric response function, the initial constant is the
center of the strip or one of its border. These special points have
$x_{gk}=\varepsilon$. For the asymmetric case, the initial
constant must be determined resorting to other properties of the
COG algorithms.

The presence of noise modifies this picture introducing an average
over the noise realization. To render the approach less heavy we
will neglect  this average, but now equation~\ref{eq:equation08} becomes the
definition of the function $\varepsilon_k (x_{g k})$.

The uniform distribution of events $\varepsilon_j$ on the array of
periodic detector generates a periodic probability distribution
$\Gamma^p(x_{gk})$ (normalized on a period), and the solution of
equation~\ref{eq:equation08}, for the symmetric configuration, is given
by:
\begin{equation}\label{eq:equation09}
\varepsilon_k(x_{gk})=-\frac{\tau}{2}+\tau\int_{-\tau/2}^{x_{gk}}\Gamma^p(x)\mathrm{d}
x
\end{equation}
The initial constant used in~\cite{belau} is
$\varepsilon_k(x_{gk}=0)=0$, but, as we discussed above,
$\varepsilon_k(x_{gk}=-\tau/2)=-\tau/2$ and
$\varepsilon_k(x_{gk}=0)=0$ are exact for symmetric $\varphi(x)$
and for symmetric response function. In all the other cases,
the required correction will be indicated with
$\Delta_{0k}$.

It is easy to show the periodicity of
$\varepsilon_k(x_{gk})-x_{gk}$, in fact, due to the periodicity
and the normalization of $\Gamma^p$ we may rewrite the
equation~\ref{eq:equation09} as:
\begin{equation}\label{eq:equation09_a}
\varepsilon_k(x_{gk})=x_{gk}+\int_{-\tau/2}^{x_{gk}}(\tau\Gamma^p(x)-1)\mathrm{d}\,x\,.
\end{equation}
The integral is a periodic function of $x_{gk}$, and we
express it as a FS:
\begin{equation}\label{eq:equation08b}
\begin{aligned}
&\varepsilon_k(x_{gk})=x_{gk}+\sum_{n=-\infty}^{+\infty}\alpha_n
\,\mathrm{e}^{(\mathrm{i} 2\pi\,n x_{gk}/\tau)}\\
&\alpha_n=\frac{1}{\tau}\int_{-\tau/2}^{+\tau/2}[\varepsilon_k(x_{gk})-x_{gk}]\, \mathrm{e}^{(-\mathrm{i}
2\pi\,n\, x_{gk}/\tau)}\mathrm{d}\,x_{gk},
\end{aligned}
\end{equation}
and with the correction $\Delta_{0k}$:
\begin{equation}\label{eq:equation09a}
\varepsilon_k(x_{gk})=x_{gk}+\sum_{n=-\infty}^{+\infty}\alpha_n
\exp(\mathrm{i} 2\pi\,n x_{gk}/\tau)+\Delta_{0k}\,.
\end{equation}
In the definition of the $\alpha_n$, the $k$-index, the number of
strips used in the algorithm, is not explicitly reported, but it is
evident that $\alpha_n$ depends from $k$.

With low noise, the function
$\varepsilon_k(x_{gk})$ is a good approximation of noiseless form,
and it sits on the most probable values of $\varepsilon$ for any $x_{gk}$.
This property is crucial for any best fit in a track reconstruction.
The absence of the correction $\Delta_{0k}$ introduces an average
systematic shift of $\varepsilon_k$ respect to the true
$\varepsilon$, quite evident in the simulations.
\subsection{Correction of the $\eta$ Algorithms}
We calculate $\Delta_{0k}$ exploring the mean value of the
differences $\varepsilon_k(j)-\varepsilon(j)$ in a case of a large number $N$ of
events and uniform distribution on a given strip as defined. The mean value must be zero in the absence of systematic errors:
\begin{equation}\label{eq:equation10_a}
\begin{aligned}
&\frac{1}{N}\sum_{j=1}^N[\varepsilon_k(j)-\varepsilon(j)]=
\frac{1}{N}\sum_{j=1}^N[x_{gk}(j)-\varepsilon(j)]\\
&+\frac{1}{N}\sum_{j=1}^N[\sum_{n=-\infty}^{+\infty}\alpha_n\, \mathrm{e}^{(\mathrm{i}
2\pi \,n x_{gk}(j)/\tau)}]+\Delta_{0k}
\end{aligned}
\end{equation}
The mean value of the
FS, weighted with the probability $\Gamma^p(x_{gk})$,
gives $\alpha_0$~\cite{landi03}. Adding and subtracting the position of the
strip with the maximum signal $\mu_j$, equation~\ref{eq:equation10_a}
becomes:
\begin{equation}\label{eq:equation011}
\frac{1}{N}\sum_{j=1}^N[\varepsilon_k(j)-\varepsilon(j)]=
\frac{1}{N}\sum_{j=1}^N[x_{gk}(j)-\mu_j]-\frac{1}{N}\sum_{j=1}^N[\varepsilon(j)-\mu_j]+\alpha_0+\Delta_{0k}
\end{equation}
The mean of $(\varepsilon(j)-\mu_j)$ is independent from the COG
algorithm, and we calculate it in the easier condition.
We use equation~\ref{eq:equation05_b} with all the signal strips in
the COG algorithm (four or five at most), to near the condition of equation~\ref{eq:equation03}. We will call
this $x_{g \infty}$, and equation~\ref{eq:equation05_b} becomes:
\begin{equation}\label{eq:equation07b}
\frac{1}{N}\sum_{j=1}^N(x_{g\infty}(j)-\mu_j)=
\frac{1}{N}\sum_{j=1}^N(\varepsilon(j)-\mu_j)+\delta_g\,.
\end{equation}
The average of $\varepsilon(j)-\mu_j$ of the unknown exact
impact points is reduced to known quantities. Substituting in
equation~\ref{eq:equation011}, and imposing the zero average of
$(\varepsilon_k(j)-\varepsilon(j))$ the equation for $\Delta_{0k}$
becomes:
\begin{equation*}
\begin{aligned}
&\frac{1}{N}\sum_{j=1}^N(x_{gk}(j)-\mu_j)-\frac{1}{N}\sum_{j=1}^N(x_{g\infty}(j)-\mu_j)+\delta_g+\alpha_0+\Delta_{0k}=0\\
&\Delta_{0k}=\frac{1}{N}\sum_{j=1}^N(x_{g\infty}(j)-\mu_j)-\frac{1}{N}\sum_{j=1}^N(x_{gk}(j)-\mu_j)
-\alpha_0-\delta_g
\end{aligned}
\end{equation*}
We have to recall that the two expressions $\sum_{j=1}^N(x_{g\infty}(j)-\mu_j)/N$ and
$\sum_{j=1}^N(x_{g\,k}(j)-\mu_j)/N$ are the averages of the COGs
calculated in our reference system of the maximum signal strip.
The constant $\alpha_0$ embodies the initial conditions (not
limited to 0 or $-\tau/2$) and the correction $\Delta_{0k}$ eliminates
any reference to the initial integration contant.

To be complete, the correction $\Delta_{0k}+\alpha_0$ eliminates even the
systematic error of the COG algorithm (with $k$-strips) due to the non zero
$\delta_g$ and to the eventual loss given by the limitation in the strip
number. The loss has rarely a significative effect, but we are able to
consider it. For the COG indicated with $x_{g\infty}$ the correction is simply
$-\delta_g$. The residual non zero average is the mean value of $(\varepsilon(j)-\mu_j)$
that the asymmetry modifies respect to its zero value in the symmetric case.

In the simulations of ref.~\cite{landi03}, the COG algorithm with
four strips was a good approximation for $x_{g\infty}$, here we will
use even the five strip algorithms.
Due the physical meaning of $\delta_g$, once its value is obtained,
the correction of any position algorithm and for any incidence angle can be
implemented. In the following we will need the correction
$\Delta_{0\infty}$ to the $\eta$-algorithm
obtained starting from $x_{g\infty}$, this correction is given by $-\alpha_0-\delta_g$.

An indication of $\delta_g\neq 0$ is given by a non
zero value of the average of $(x_{g\infty}(j)-\mu_j)$ for the
orthogonal incidence. This averages can not be zero for equation~\ref{eq:equation07b}.
It is a sum of two unknown quantities, and another equation is necessary to extract their values.
To estimate  $\delta_g$, we need to
reconstruct $\varphi(\varepsilon)$ and explore its asymmetry.

\section{ Determination of $\delta_g$}
\subsection{Signal reconstruction}

In refs.~\cite{landi01,landi03} we demonstrated an equation to obtain the signal
distribution from the COG algorithm. The $\varphi(\varepsilon)$ is given by:
\begin{equation}\label{eq:equation10}
\frac{\mathrm{d} x_g(\varepsilon)}{\mathrm{d}\varepsilon}=
\tau\sum_{n\in\,\mathbb{Z}}\varphi(n\tau-\varepsilon-\tau/2)\,.
\end{equation}
In the derivation of equation~\ref{eq:equation10} the response function is assumed to be the lossless interval function, and $\varphi(x)$ is the signal distribution. The
expression~\ref{eq:equation10} is the sum of copies of
$\varphi(\tau/2-\varepsilon)$ shifted of $n\tau$ with $n\in
\mathbb{Z}$. This gives a periodic function with overlaps of the tails
(aliasing) if the range of $\varphi(x)$ is greater than $\tau$.
For ranges less than $\tau$ the reconstruction is faithful. If the
range of $\varphi(x)$ is greater than $\tau$, the assembly of a
set of contiguous interval functions avoids this limitation.

For fluctuating signal distributions, as in the case of a MIP,
equation~\ref{eq:equation10} defines an average signal distribution.

\subsection{Generic Response Function}

The reconstruction of $\varphi(x)$ of equation~\ref{eq:equation10} requires
the response function as a pure interval function of size
$\tau$, rarely this condition is verified, and a generic response
function  produces a redefinition of $\varphi(x)$.
If the lossless condition
equation~\ref{eq:equation03} is maintained, we proved in
ref.~\cite{landi01} that the response function must be a
convolution of an interval function with another (arbitrary and
eventually asymmetric) function g$_1(x)$. In this case,
the function of
Eq.~\ref{eq:equation10} is
the convolution of the true signal distribution with g$_1(x)$.
We will continue to call $\varphi(x)$ any result of $\mathrm{d} x_g(\varepsilon)/\mathrm{d}\varepsilon$
even if it deviates from the true signal distribution.

\begin{figure}[h!]
\begin{center}
\includegraphics[scale=0.85]{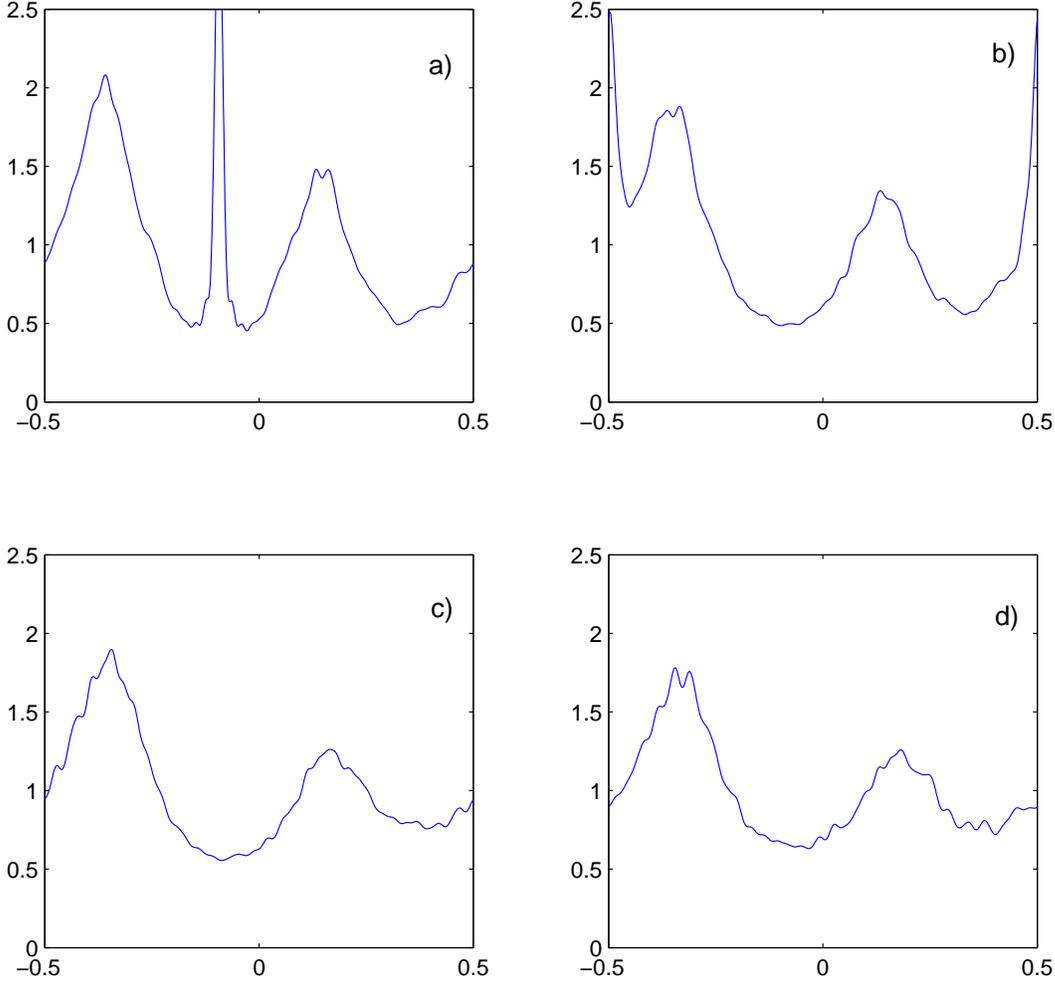}
\caption{\em Reconstruction of $\varphi(x_k)$ (with noise) using different
numbers of strips: a) with 2 strips, b) with 3 strips, c) with 4
strips, d) with 5 strips.  The asymmetry parameter is 0.07 and 45000 events.}
\label{fig:figura03_a}
\end{center}
\end{figure}

In ref.~\cite{landi03}, we explored a possible form of the
response function for microstrip detector with floating strips,
and the following form reproduces the main aspects of the data:
\begin{equation}\label{eq:equation21}
\begin{aligned}
p(x)=\int_{-\infty}^{+\infty}\Pi(x-x')\big\{&0.45\;[\delta
 (x'-1/4)
 +\;\delta(x'+1/4)]\\
&+0.05[\;\delta(x'-1/2)+\;\delta(x'+1/2)]\big\}\,.
\end{aligned}
\end{equation}
(This form is surprisingly similar to that measured in ref.~\cite{infrared_laser}.)
Here, the reconstruction of equation~\ref{eq:equation10} generates the
convolution of the signal distribution with the  four Dirac
$\delta$-functions of equation~\ref{eq:equation21}, the low intensity
Dirac $\delta$-functions have a negligible effect,
but the effects of the two main $\delta$'s are
clearly seen in figure~\ref{fig:figura03_a} as two copies of the
signal distribution.

If the response function is asymmetric, the asymmetry is
contained in $x_g(\varepsilon)$ and transferred to the
reconstructed function. The asymmetry
transfer to $\varphi(x)$ does not allow a direct extraction of $\delta_g$.
We have to resort to an indirect procedure.

\subsection{Inverse function of $\varepsilon_k(x_{gk})$}

The form of equation~\ref{eq:equation08b} for $\varepsilon_k(x_{gk})$
is not well suited for our needs. Its inverse function
$x_{gk}(\varepsilon_k)$ is of better use, and it is
expressed by:
\begin{equation}\label{eq:equation22}
x_{gk}(\varepsilon_k)=\varepsilon_k+\sum_{n=-L}^{+L}\beta_n\exp(\mathrm{i}\frac{2
\pi n}{\tau}\varepsilon_k)
\end{equation}
\[\beta_n=\frac{1}{\tau}\int_{-\tau/2}^{\tau/2}\big[x_{gk}-\varepsilon_k(x_{gk})\big]\exp\big(-\mathrm{i}\frac{2
\pi n}{\tau}\varepsilon_k(x_{gk})\big)\Gamma^p(x_{gk})\mathrm{d} x_{gk}\]
where $\varepsilon_k(x_{gk})$ is the result of
equation~\ref{eq:equation09} and $L$ the maximum wave-number used, $L$ around 45 is a reasonable
cut off even if the limits $\pm\infty$ will be often used. All the forms of $\varphi(x)$ are obtained
differentiating equation~\ref{eq:equation22} respect to $\varepsilon_k$.

As discussed, deformations are introduced in $\varphi(x)$
by the differences of $p(x)$ respect to an interval function.
Another set of important deformations are produced by the loss.
Two type of loss are encountered: the intrinsic loss of the strip
and the loss given by the suppression of non-zero signal strips.
The first type of loss operates as an additional smooth deviation of the
response function from the pure interval function, it has a negligible effect
on our procedures.
The second type of loss introduces a strong deformations in $\varphi(x)$.
The presence of any type of loss is explicitly excluded by the form of
equation~\ref{eq:equation10}, but we can, in any case, differentiate
equation~\ref{eq:equation22} and explore its results. In the absence of
noise, the deformations given by the second type of loss assume the form of Dirac
$\delta$-functions. The limitation in the number of terms in equation~\ref{eq:equation22}
gives finite peaks. The exclusion of signal strips in the COG algorithm generates
forbidden $x_g$-values. Here the probability $\Gamma^p$
is zero and $\varepsilon(x_g)$ has an interval of constant value. If we
insist to invert the function $\varepsilon(x_g)$ this constant
horizontal segment becomes a vertical segment, and the differentiation
generates a Dirac $\delta$-function. In general, if the strip number is even, one
aspects peaks around $\varepsilon_k\approx 0$, for odd strip
numbers the peaks are for $\varepsilon_k\approx \pm \tau/2$. The amplitudes of the peaks are
proportional to amplitude of the signal function acquired by the
excluded strips~\cite{landi01}.  For
$k=2,3$ clear peaks are present in the
reconstructions of ref.~\cite{landi03} and figure~\ref{fig:figura03_a}.
For $\delta_g\neq 0$ and
$k=2$, the peak is not in zero due to the $-1/2$ as the lower
integration limit of equation~\ref{eq:equation09},  this fixes  $\varepsilon_k=-1/2$
to coincide with $x_{gk}=-1/2$. In fact, the peaks of $x_{g3}$ are at
$\varepsilon_3=\pm 1/2$. With the lower integration limit to zero,
$\varepsilon_k=0$ coincides with $x_{gk}=0$ and the peak for
$k=2$ would be at $\varepsilon_2=0$. It is evident that with an asymmetry
in the response function nor $x_{gk}=-1/2$ nor $x_{gk}=0$ are exact, and
$\Delta_{0k}$ fixes the correct $\varepsilon_k$ and the correct positions
of the peaks.
The almost total suppression of the loss eliminates the peaks for $k=4$ and $k=5$ in
figure~\ref{fig:figura03_a}.

A due care must be devoted to avoid numerical instabilities.
Intervals where equation~\ref{eq:equation22} does not exist (due to
zero values of $\Gamma^p$) are easily encountered in noiseless
case. The noise helps to avoid $\Gamma^p=0$, but it easily adds
other unwanted artifacts. The Cesaro's method
of arithmetic means~\cite{libroFT2} attenuates some numerical instabilities.

\subsection{Analytical form of $x_g(\varepsilon)$}

The exploration of the analytical form of $x_g(\varepsilon)$ clarifies
our path toward $\delta_g$. Here all the properties
of the detector and the signal distribution are explicitly underlined.
In the case of orthogonal incidence, the incoming
signal distribution is symmetric and its FT $\Phi(\omega)$ is real and
symmetric. The response function p$(x)$ has asymmetries and its FT
$P(\omega)$ is complex with $P(-\omega)=P^*(\omega)$. The form of
$x_g(\varepsilon)$, with $\delta_g$ the first momentum of $p(x)$,
is~\cite{landi03} (with $\tau=1$):
\begin{equation}\label{eq:equation23}
x_{g}(\varepsilon)=\varepsilon+\delta_g+i \sum_{k\neq 0,
k=-\infty}^{+\infty}\Phi(-2k\pi)\mathrm{P}'(-2k\pi) \exp(i\,
2k\pi\varepsilon)\,,
\end{equation}
where $\mathrm{P}'(\omega)$ is the first derivative of
$\mathrm{P}(\omega)$ respect to $\omega$, and $\varepsilon$ is the
impact point. We know that, even in the best condition,
the equation~\ref{eq:equation09} for $\varepsilon_k$  has
an incorrect initial constant. To handle the asymmetric case, we have to
generalized a new $\overline{\varepsilon}_k(x_{gk})$ defined for any initial
condition $x_{gk}^0$ beyond  the $x_{gk}^0=-1/2$ of equation~\ref{eq:equation09}:
\begin{equation*}
\overline{\varepsilon}_k(x_{gk})=x_{gk}^0+\int_{x_{gk}^0}^{x_{gk}}\Gamma(x)^p\mathrm{d} x
\end{equation*}
The correction procedure must work for any $x_{gk}^0$.
It is evident that $x_{gk}^0=-1/2$ or $x_{gk}^0=0$ have
the minimal corrections being exact in the symmetric case.
The constant $\varepsilon(x_{gk}^0)-x_{gk}^0$ is now the
difference of $\overline{\varepsilon}_k$ from $\varepsilon$,
given the initial constant $x_{gk}^0$.
Substituting $\varepsilon$ with
$\overline{\varepsilon}_k$ in equation~\ref{eq:equation23}, we have:
\[\varepsilon=\overline{\varepsilon}_k+(\varepsilon(x_{gk}^0)-x_{gk}^0)\]
\begin{equation}\label{eq:equation25}
x_{g}(\overline{\varepsilon}_k)=\overline{\varepsilon}_k+(\varepsilon(x_{gk}^0)-x_{gk}^0)+\delta_g+
\sum_{n\neq 0, n=-\infty}^{+\infty}\widetilde{\beta_n} \exp[i\,
2n\pi(\overline{\varepsilon}_k+\varepsilon(x_{gk}^0)-x_{gk}^0)]\,.
\end{equation}
Remembering equation~\ref{eq:equation22}, the comparison with equation~\ref{eq:equation23}
 gives:
\begin{equation}\label{eq:equation26}
\begin{aligned}
&\beta_n=\widetilde{\beta_n} \exp[i\,
2n\pi(\varepsilon(x_{gk}^0)-x_{gk}^0)]\ \ \ n\neq 0\\
&\beta_0=(\varepsilon(x_{gk}^0)-x_{gk}^0)+\delta_g\\
&\widetilde{\beta_n}=i\,\Phi(-2n\pi)\mathrm{P}'(-2n\pi)
\end{aligned}
\end{equation}
as expected
$\beta_0$ is the sum of the two unknown $\delta_g$
and the shift
$(\varepsilon(x_{gk}^0)-x_{gk}^0)$. To extract $\delta_g$ we need
another equation. The derivative $\mathrm{d}
x_g(\overline{\varepsilon}_k)/\mathrm{d}\overline{\varepsilon}_k$ does not
contain $\beta_0$, it has a shift of
$(\varepsilon(x_{gk}^0)-x_{gk}^0)$ respect to the differentiation in
the exact $\varepsilon$. This shift is present as a phase factor
in equation~\ref{eq:equation26}, and it goes to increase the asymmetry
of $\varphi(\overline{\varepsilon}_k)$. In the symmetric
configuration the phase relations are easy: all the $\Phi(-2n\pi)\mathrm{P}'(-2n\pi)$ are real and $\widetilde{\beta_n}$  imaginary.

Due to the special form of the of the relation of $\beta_n$
and $\widetilde{\beta_n}$, we can add a fictitious phase
parameter $2\pi n\xi$ to $\beta_n$ to modify the asymmetry of $\varphi(\overline{\varepsilon}_k)$. For small
asymmetry, we expect that this asymmetry variation of $\varphi(\overline{\varepsilon}_k)$ reaches its minimum when all the $\beta_n$ coincide with $\widetilde{\beta_n}$. The phase factors of the $\widetilde{\beta_n}$ are given by the intrinsic asymmetry of the response function, and are non trivial functions of $n$. These phase relations are not eliminated by the trivial transform  implied by a  global shift of $\varphi$ and the  intrinsic asymmetry cannot be reduced. An asymmetry parameter with a
small sensitivity to the noise is:

\begin{equation}\label{eq:equation26_a}
\Omega(\xi)=\int_{-1/2}^{1/2}
\big[\varphi(\xi+\overline{\varepsilon}_k)-\varphi(\xi-\overline{\varepsilon}_k)\big]^2\mathrm{d}
\overline{\varepsilon}_k\,.
\end{equation}

Here $\varphi(\overline{\varepsilon}_k)$ is $\mathrm{d}
x_g(\overline{\varepsilon}_k)/\mathrm{d} \overline{\varepsilon}_k$.
The minimum of $\Omega(\xi)$ is obtained for a $\xi_m$ given by:
\[ \xi_m=-(\varepsilon(x_{gk}^0)-x_{gk}^0)\,, \]
this is the second equation that allows the use of
$\beta_0$ in equation~\ref{eq:equation26} to extract $\delta_g$:
\begin{equation}\label{eq:equation27}
\delta_g=\beta_0+\xi_m\,.
\end{equation}
$\Omega(\xi)$ is expressed with $\beta_n$ as:
\begin{equation}\label{eq:equation27a}
\Omega(\xi)=2\sum_{n\in
\mathbb{Z}}\big[|\beta_n|^2+\beta_n^2\exp(\mathrm{i} 4\pi
n\xi)\big](2\pi n)^2\,.
\end{equation}
The first term is a constant and the $\xi$-dependence is a
periodic function of period $1/2$. When $\delta_g=0$ and $x_{gk}^0=0$ or $-1/2$
and $\beta_n=\widetilde{\beta}_n$ it is easy to verify the
minimum for $\xi=0$ given that $\widetilde{\beta_n}^2=-|\beta_n|^2$. In general, the minima of equation~\ref{eq:equation27a} produce the corrections of the $\eta$-algorithm for all
the initial $x_{gk}^0$.

\begin{figure}[h!]
\begin{center}
\includegraphics[scale=0.9]{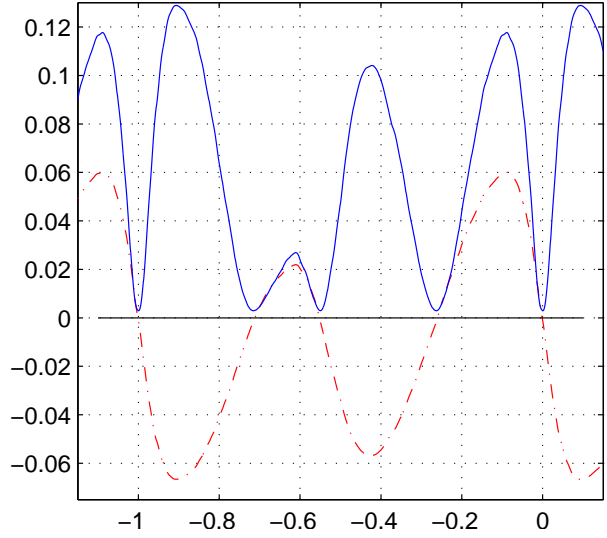}
\caption{\em Continuous line (blue): Asymmetry $\Omega(0)$ calculated for all the initial conditions $x_{g4}^0$ from -1 to 0. The dash-dotted line (red) is the correction $\varepsilon_4^0-x_{g4}^0$, the asymmetry has minima when $\varepsilon_4^0-x_{g4}^0=0$. Noiseless simulation with $\zeta=0.02$ and floating strip sensor}
\label{fig:figura05_b}
\end{center}
\end{figure}

To see the effectiveness of the minimization of equation~\ref{eq:equation27a}, we calculated $\Omega(\xi)$ with $\xi=0$ for all the initial conditions $x_{gk}^0$ from $-1$ and
$0$. In this case $\Omega(0)$ has minima where $\beta_n=\widetilde{\beta}_n$
or, more precisely, in the points where $\varepsilon_k^0-x_{gk}^0=0$. In figure~\ref{fig:figura05_b} we report $\Omega(0)$ in function of the initial $x_{gk}^0$ and effectively it has evident minima when $\varepsilon_k^0-x_{gk}^0=0$. In the simulations we use a low asymmetry $\zeta=0.02$, and a noiseless simulation and $x_{g4}$ algorithm for a floating strip sensor. Figure~\ref{fig:figura05_b} gives an empirical support to our research of the minima for $\Omega(\xi)$.

In general, any initial condition can be used, but, the $\xi_m$ are
widely different with an inefficient minimum search.
The special form of ($\beta_0$) and the following substitution
simplifies the search and eliminates the explicit dependence
from the initial conditions:
\begin{equation}\label{eq:equation28}
\begin{aligned}
&\beta_n'=\beta_n\exp(-\mathrm{i}\,2\pi n
\beta_0)\\
&\beta_n'=\widetilde{\beta}_n \exp(-\mathrm{i}\,2\pi n
\delta_g)\,,
\end{aligned}
\end{equation}
and, for any $x_{gk}^0$, equation~\ref{eq:equation27} is reduced to:
\begin{equation*}
\delta_g=\xi_m\,.
\end{equation*}

A presence of a small loss has a negligible effect on this approach.
Large loss, signaled by the presence of peaks around zero or $\pm1/2$,
can strongly modify the minimum search.For example, in our first set of
simulations, $x_{g2}$ and $x_{g3}$ have minima very different
from that of $x_{g4}$, $x_{g5}$.

\subsection{Simulations}

The simulated data are generated as discussed in
ref.~\cite{landi03}. For the floating strip case, we modify the response function
breaking the symmetry of the two most important
Dirac-$\delta$ functions of equation~\ref{eq:equation21}. We add to the
first Dirac-$\delta$ function a constant $\zeta$ and a $\zeta$ is
subtracted to the other one to save the normalization, figure~\ref{fig:figura03_a}
has $\zeta=0.07$ . This type of asymmetry
looks similar to the one observed in the test-beam data, but we amplify
the effect. In any
case, this is only a numerical experiment to see the efficiency of
the $\delta_g$ determination. We will compare with the noiseless
simulation to see the effect of the noise.
\begin{figure}
\begin{center}
\includegraphics[scale=0.9]{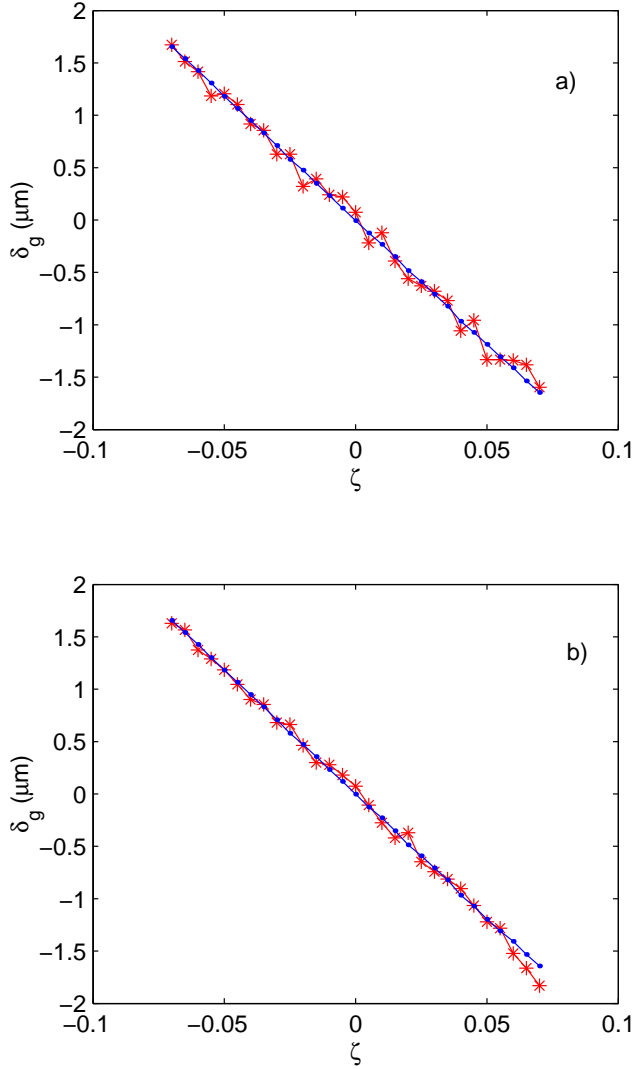}
\caption{\em Noiseless case, the blue dots are the results of equation 2.14, 
the red asterisks are the $\delta_g$ obtained by the minima of equation 3.9. 
The plot $a$ is given by $x_g$ with four strips and $b$ with five strips.   }
\label{fig:figura06_a}
\end{center}
\end{figure}

Figure~\ref{fig:figura06_a} shows the determination of
$\delta_g$ with the procedure illustrated above. The simulated data
are noiseless, but even here we see fluctuations of $\delta_g$
from equation~\ref{eq:equation07b}. The fluctuations
originate from the reconstruction that requires the extraction of
$\varphi(\varepsilon)$ from histograms and, due to a finite set of data
(45000 events), the procedure adds an effective noise that is  lower in the
case of $\delta_g$ calculated with five
strips. Here the attenuation of the fluctuations
could be due a reduction of the slight loss, that is present in the four strip simulation
due to the suppression
of the signal (convolution of gaussians~\cite{landi03}) collected by the fifth strip.
This loss is too low to produce a peak, but it contributes to the effective
noise of $\Gamma^p$.

The realistic case (with noise) fluctuates more than the noiseless
case. Even here the $\delta_g$ calculated with five strips has
less fluctuations than that calculated with four strips. The RMS
error is 0.1 $\mu$m for the five strips case and 0.2
$\mu$m for the four strips case.
\begin{figure}[h!]
\begin{center}
\includegraphics[scale=0.9]{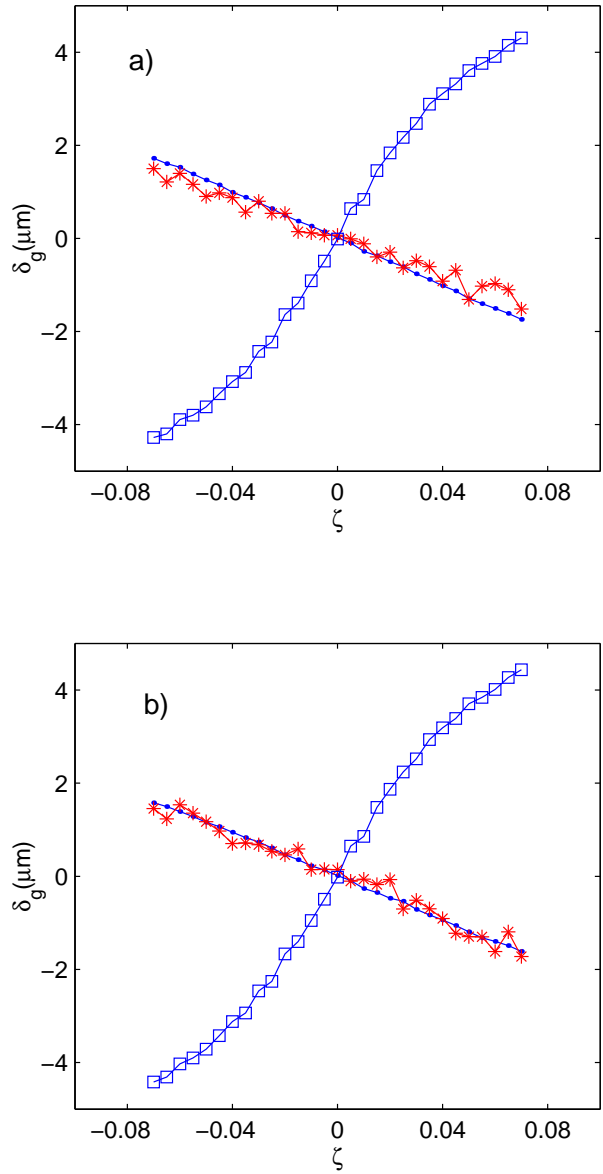}
\caption{\em Noisy case. The dots are the results of
equation 2.14.
The asterisks are the $\delta_g$
given by the minima of equation 3.9,
and the
squares are the COG averages. Plot $a)$ is for the four
strip $x_g$ and $b)$ is for the five strip $x_g$. }
\label{fig:figura07_a}
\end{center}
\end{figure}
In figure~\ref{fig:figura07_a} we reported the averages
$\sum_{j=1}^N(x_{g\,k}(j)-\mu_j)/N$ that is the signal of a non
zero $\delta_g$. The form of asymmetry generation, we used, gives an
amplification of the COG averages by (relatively) small $\delta_g$.
For the floating strip side, the introduction of
$\delta_g$ could be a minor correction around half micron, with all the
sensors oriented identically a parallel shift of the track is implied.
If some sensor has a reverse orientation $\delta_g$ change sign and gaps of
a micron are present in the tracks. Even if these constant shifts could be
corrected by the alignment procedures, it is a good practice to have estimators
free of bias when possible.

\subsection{Normal strips}

We explore the strategy of the extraction of $\delta_g$ for the
case of "normal" strips. Even now the impact direction is
orthogonal to the detector plane. At this angle, the detector
resolution is low due to the concentration of a large parte of the
signal in a single strip. To be consistent with the real detector,
the simulated noise is doubled respect
to the case of the floating strip sensor, and its effect strongly
deteriorates the extraction of $\delta_g$.
Here we have no indications of the type of asymmetry, the mean value
of $x_{g4}$ is different from zero, but the reconstruction does not
show evident asymmetry.
We produce the asymmetry with an additional
Dirac-$\delta$ function
$\zeta\,\delta(x-\tau/4)$ convoluted with the usual interval
function to have an everywhere flat efficiency. The
values of $\zeta$ are all positive, we must avoid
negative values of the response function and of $\mathrm{d}
x_g(\varepsilon)/\mathrm{d}\varepsilon$.

For the noiseless case, the reproduction of $\delta_g$ is
reasonable for all the $\zeta$ values even if the fluctuations
introduced by the finite number of events is higher than the
corresponding case of the floating strips. The addition of the noise
changes drastically the results, the determination of $\delta_g$
degrades rapidly at increasing $\zeta$, now large values of
$\delta_g$ are connected to lower values of
$\sum_{j=1}^N(x_{g\,k}(j)-\mu_j)/N$. Here we report even the results of the two-strip algorithms, and they are better than the four strip case. In general, the loss of the two strip COG could give incorrect results, the peak around zero can be very high and it drastically deforms $\varphi$.  In this case, the noise washes away the peaks, and the noise reduction of the two strip algorithm gives a $\varphi(\varepsilon_2)$ more sensible to the asymmetry parameter
$\delta_g$ than $\varphi(\varepsilon_4)$. The use of the two strip algorithms could be interesting in presence of high noise.

\begin{figure}[h!]
\begin{center}
\includegraphics[scale=0.9]{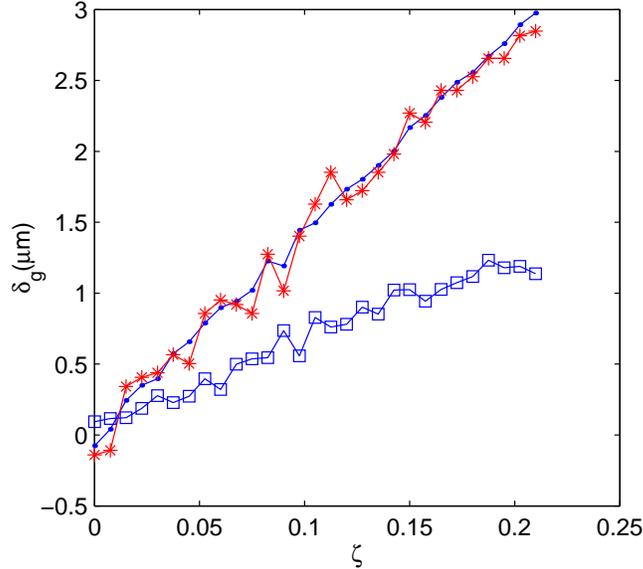}
\caption{\em Noiseless case. The blue dots indicate the results of
equation 2.14.
The red asterisks are the $\delta_g$
obtained by the minima of equation 3.9 
with four-strip algorithm. The blue squares are the COG averages}
\label{fig:figura09_a}
\end{center}
\end{figure}
\begin{figure}[h!]
\begin{center}
\includegraphics[scale=0.9]{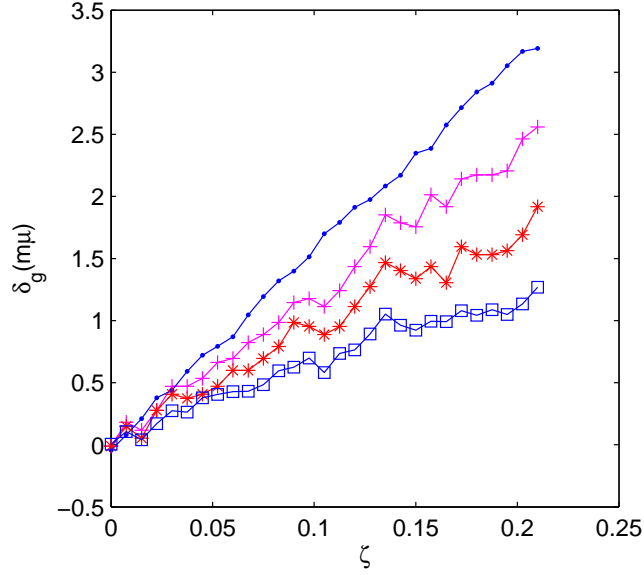}
\caption{\em The effect of the noise. The meanings of the symbols
are that of figure 5.
The magenta crosses indicate  the $\delta_g$  obtained with the two-strip algorithm. }
\label{fig:figura09_b}
\end{center}
\end{figure}
\section{Non orthogonal incidence and Lorentz angle}

\subsection{Asymmetry due to non orthogonal incidence}

In the previous calculations, the orthogonality of the incoming particles was
often recalled as a fundamental condition to access to the asymmetry of the
response function. But, the effects of the deviations from the orthogonality
must be explored to test the robustness of the algorithm.
The $\beta_n$ of equation~\ref{eq:equation27a} have terms $\Phi(-2\pi n)$ (FT of the true incoming $\varphi(x)$) in their definition~\ref{eq:equation26}, for the orthogonal incidence any $\Phi(-2\pi n)$ is real (and symmetric).  An angular deviation $\theta$ ($\theta=0$ for orthogonal incidence) from zero adds phase factors to the $\Phi(-2\pi n)$ and it introduces a large asymmetry in
equation~\ref{eq:equation27a}. Some plots of $\varphi(x)$ with $\theta\neq 0^o$ are
reported in ref.~\cite{landi03}. For example, a value of $\theta=0.2^o$ gives $\xi_m=0.3\  \mu m$ for a symmetric response
function (floating strip case). Thus, the asymmetry of the signal
distribution, can easily mask the asymmetry of the response function.
The $\theta$ data must have sufficient accuracy to detect small effects.
In any case, the asymmetry induced by $\theta\neq 0$ changes its sign with the sign
of $\theta$ and $\delta_g$ remains constant. So, for sufficiently small angles,
where the total asymmetry is almost linear, the collection of data at various angles
around $\theta=0$ and the fit to the corresponding $\xi_m$ with a low degree polynomial
function can give a better value of $\delta_g$.

\begin{figure}[h!]
\begin{center}
\includegraphics[scale=0.9]{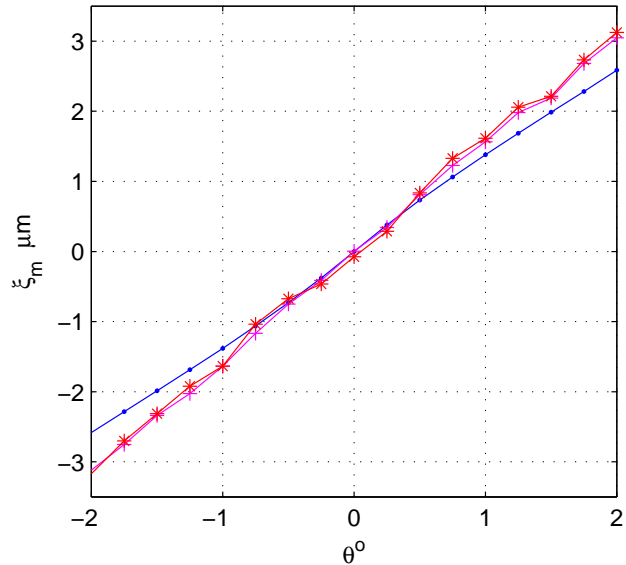}
\caption{\em  Asymmetry $\zeta=0$, floating strip sensors. Dotted line (blue) $\xi_m$ on the convolution  $(g_1*\varphi)$ of the model signal distribution,  asterisk line (red) $\xi_m$ of $\varphi(\overline{\varepsilon}_4)$ and crosses line (magenta) is the $\xi_m$ for the noiseless case.} \label{fig:figura09_c}
\end{center}
\end{figure}

To explore the variation of $\xi_m$ from $\theta$, we process the convolution ($\varphi*g_1$) of
our model~\cite{landi03} of $\varphi(x)$, with the machinery of equation~\ref{eq:equation26_a}.
This is a very easy operation due the explicit FT expression of ref.~\cite{landi03}. The results
are illustrated in figure~\ref{fig:figura09_c} for the $g_1$ of floating strip sensors. A good
linear relation is obtained for small $\theta$ values, this linearity is driven by two effects:
a phase factor proportional to the angle in the model function, and the two copies of $\varphi$
given by the two Dirac-delta of $g_1$. For comparison, in the normal strip case the absence of
the two delta adds non linear distortions. This linearity of $\xi_m$ is saved (with a small
reduction of the slope) in our reconstructed $\varphi(\overline{\varepsilon})$ and it is almost
insensible to the noise.

\begin{figure}[h!]
\begin{center}
\includegraphics[scale=0.9]{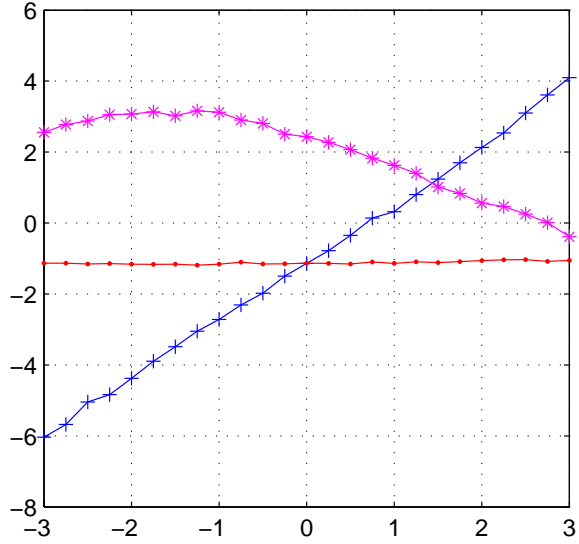}
\caption{\em Asymmetry $\zeta=0.04$, floating strip sensor. Crosses line (blue) $\xi_m$ of $\varphi(\overline{\varepsilon}_4)$,  dotted line (red) systematic error of $\varepsilon_2$ without the correction $\delta_g$ and asterisks line (magenta) is the average of $x_{g4}$. }
\label{fig:figura09_d}
\end{center}
\end{figure}

The direct application of equation~\ref{eq:equation26_a} on $\varphi*g_1$ gives $\xi_m$-values that
depend very weakly on the asymmetry $\zeta$ due an almost complete cancelation of the first
order terms. In any case, $\varphi*g_1$ is accessible only in the simulations and this cancelation is
irrelevant in the data. On the contrary, the $\eta$-algorithms introduce phase factors proportional
to $\delta_g$ in the FS-amplitudes of $\varphi(\overline{\varepsilon})$ as a global shift of the function. Thus, $\Omega(\xi)$ and equation~\ref{eq:equation28} allow the extraction of $\delta_g$ from the data. After the
proper $\delta_g$-correction of the $\eta$-algorithm, $\Omega(\xi)$ gives a minimum for $\xi_m\approx 0$.
The simulations with an asymmetry $\zeta=0.04$  are reported in figure~\ref{fig:figura09_d} at different
angles $\theta$ (step $0.25^o$) of a floating strip sensor. As in figure~\ref{fig:figura09_c}, the $\xi_m$-values have a linear relation with the angles as in the symmetric case, but the line is shifted by $-1.14 \mu m$ that is its crossing with the $\theta=0$ line. This value is the systematic error of  $\varepsilon_2(x_{g2})$
corrected with $\Delta_{02}$ for a symmetric response function, and, as expected, is constant in $\theta$ and equal to $\delta_g$. The addition of the correction $-\delta_g$ to $\Delta_{02}$ completely eliminates the systematic error in  $\varepsilon_2(x_{g2})$.  The average of $x_{g4}$ is different
from zero at $\theta=0$, signaling the asymmetry of the response function.

Similar results can be obtained for the normal strip case. The absence of the floating strip and the
high noise make the plots of $\xi_m$ to deviate from the good linearity of the floating case. Or better,
the linear approximation has a restricted  range of validity.  As in figure~\ref{fig:figura09_b}, the
asymmetry obtained from $\xi_m$ is less than the right one and part of the systematic effect remains uncorrected.
\begin{figure}[h!]
\begin{center}
\includegraphics[scale=0.9]{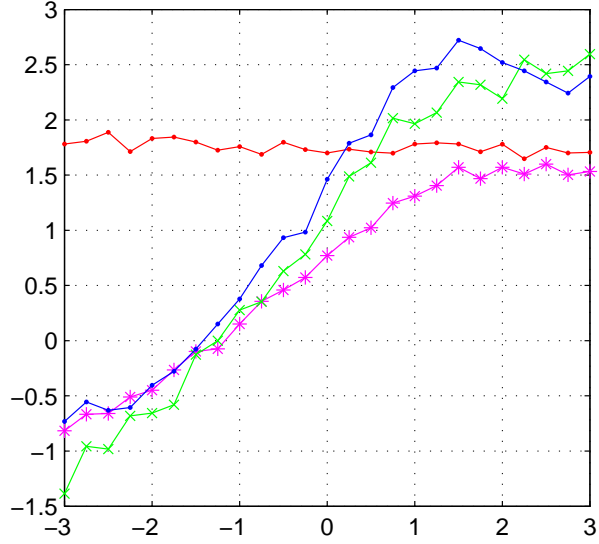}
\caption{\em Asymmetry $\zeta=0.12$, normal strip sensor. Dot-blue line  $\xi_m$ of $\varphi(\overline{\varepsilon}_2)$, cross-green line  $\xi_m$ for $\varphi(\overline{\varepsilon}_4)$.  Dot-Red line, systematic error of $\varepsilon_2$ without the correction $\delta_g$,   and asterisks-magenta line is the average of $x_{g4}$. }
\label{fig:figura10_a}
\end{center}
\end{figure}
The $\xi_m$ of $\varphi(\overline{\varepsilon}_2)$ gives a better estimation of $\delta_g$ than that
of $\varphi(\overline{\varepsilon}_4)$, but it has a strong deviation from linearity. A fit with a
low degree (3,4) polynomial function could be used. In any case, it is under study a more refined
extraction of $\varphi(\overline{\varepsilon})$ from the data with a strong suppression of the noise
distortion. Preliminary results~\cite{landi04} support a drastic improvement of the method
\subsection{Lorentz angle}

The effect of the magnetic field on the particle-holes drift in a silicon
detector is usually parameterized
as a rotation of the particle path of an angle $\theta_L$.
The rotation is around an axis parallel to field containing
the impact point. The effective COG of the track is shifted
from the true one if the strip direction is non orthogonal to
the field. The strips of the floating strip side of the PAMELA
detector are parallel to the magnetic field, and the assumed
value of $\theta_L$ is $0.7^o$. If the magnetic field has an effect
similar to a rotation on the signal distribution, the present approach
naturally measure $\theta_L$.
Usually, this measure is performed on the average
length of the clusters produced by the MIP at various incidence
angles. The minimum of the cluster size is at an incidence angle
of $-\theta_L$ (in the geometry of ref.~\cite{landi03}  where the
impact point is always in the collection plane). At this angle, the
apparent signal distribution is probably similar to that of an
orthogonal incidence, or in any case it obtains its maximal symmetry.

The method to measure $\theta_L$ with the average cluster
size has a low sensitivity just around the Lorentz angle. The data
reported in ref.~\cite{CMS} shows clearly this limitation.
It would be better to have a method with an high sensitivity just around $\theta_L$.
Our averages of $x_{g4,5}$ and $\xi_m$ have the property to go to zero
at $\theta=0$ if the signal distribution and the response function are
symmetric. Hence, in the case of symmetric condition around  $\theta=-\theta_L$,
the averages of $x_{g4,5}$ and $\xi_m$ are able to measure $\theta_L$.
With the definition of $\theta_{eff}$:
\begin{equation*}
    \tan(\theta_{eff})=\tan(\theta)+\tan(\theta_L),
\end{equation*}
the COG algorithm sees a particle track with $\theta_{eff}$ bending angle,
and its reconstruction has an effective shift of the true COG of:
\begin{equation*}
    \Delta_L=\frac{d}{2}\tan(\theta_L),
\end{equation*}
with $d$ the depletion length of the detector ($300$ $\mu m$ in our case of completely depleted sensors).
The correction $\Delta_L$ must be subtracted by any reconstruction algorithm.
\begin{figure}[h!]
\begin{center}
\includegraphics[scale=0.9]{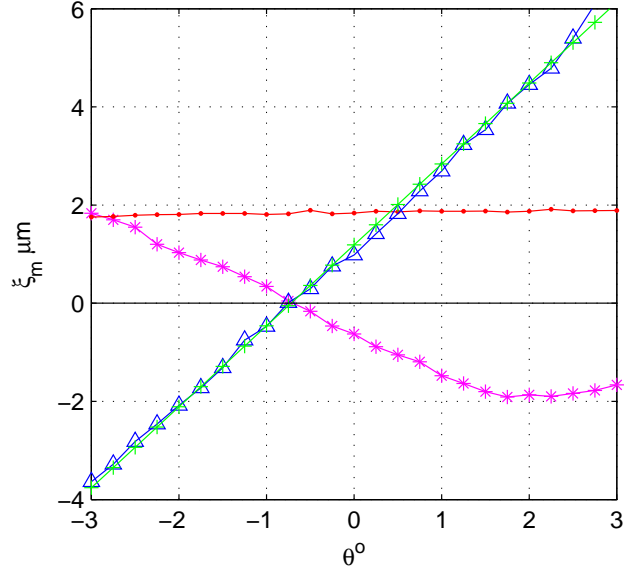}
\caption{\em Lorentz angle $\theta_L=0.7^o$ symmetric floating strip sensor. Triangles-blue line $\xi_m$
of $\varphi(\overline{\varepsilon}_4)$. Crossed-green line: linear interpolation of $\xi_m$ .
Dot-red line  systematic error of $\varepsilon_2$ without the correction $\Delta_L$,
and asterisks-magenta line is the average of $x_{g4}$.}
\label{fig:figura10_b}
\end{center}
\end{figure}

Figure~\ref{fig:figura10_b}  illustrates the sensitivity of
$\xi_m$ and the of average of $x_{g4}$ to $\theta_L\neq 0$,
each one crosses the $\theta=0$ line around  $-0.7^o$ ($\xi_m$
at $-0.72^o$ and $x_{g4}$ at $-0.70^o$). Here the detector is
perfectly symmetric, thus, the average of  $x_{g4}$ is an easy
and sensible tool to extract $\theta_L$. The asymmetry parameter
$\xi_m$ is equally sensible, but more complex to calculate. It is
clear that the symmetry condition can be verified in the detector
without magnetic field, and the average of $x_{g4}$ must be zero
for $\theta=0$.
In the case of an asymmetric response function one has to resort to
$\xi_m$. For its structure $\xi_m$ is produced by two independent
effects: the asymmetry of the response function and the effective
deviation from orthogonality. The asymmetry of the response function
$\delta_g$ must be measured without the magnetic field and the
$x_{g4,5}$ corrected accordingly. With the corrected $x_{g4,5}$,
$\xi_m$ goes again to zero for $\theta=0$. The correction $\delta_g$
is constant with $\theta$, thus, the addition of the magnetic field
gives $\xi_m=0$ at $\theta=-\theta_L$. The presence of the asymmetry
$\delta_g\neq 0$ drastically modifies the averages of $x_{g4}$ or
$x_{g5}$, and they never go to zero for $\theta=0$, or $\theta=-\theta_L$
with the magnetic field and are not usable to measure $\theta_L$.

\begin{figure}[h!]
\begin{center}
\includegraphics[scale=0.9]{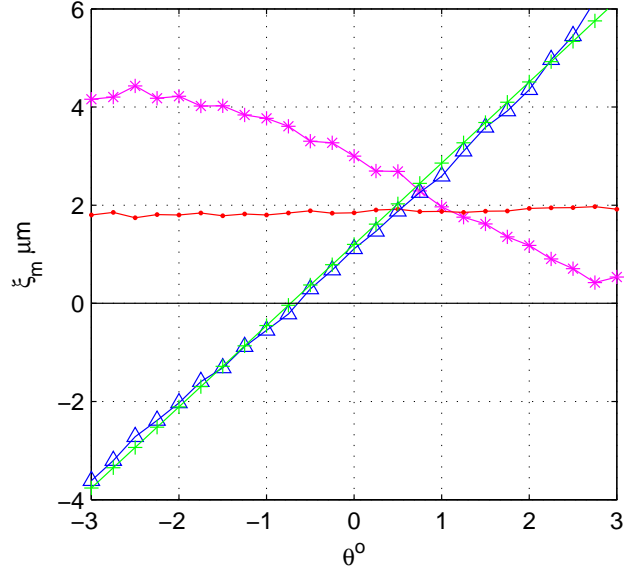}
\caption{\em Lorentz angle $\theta_L=0.7^o$ asymmetric floating strip sensor
$\zeta=0.04$. Triangles-blue line  $\xi_m$ of $\varphi(\overline{\varepsilon}_4)$.
Cross-green line: linear interpolation of $\xi_m$ .  Dot-red line systematic
error of $\varepsilon_2$ without the correction $\Delta_L$,   and asterisks-magenta line
 is the average of $x_{g4}-\delta_g$.}
\label{fig:figura10_c}
\end{center}
\end{figure}

The combined effect of the Lorentz angle and the asymmetry $\delta_g$ is
illustrated in figure~\ref{fig:figura10_c}. Here, a simulation of the
floating strip sensor with the
asymmetry of figure~\ref{fig:figura09_d} and a $\theta_L$ rotation,
is elaborate as in figure~\ref{fig:figura10_b}.

Now, figure~\ref{fig:figura10_c} shows that $\xi_m$, corrected with $\delta_g$  continues
to cross the zero line at $\theta_L=-0.7^o$.
The corrected average of $x_{g4}$ does not cross the zero line at $\theta_L=-0.7^o$ and
its use as an estimator of $\theta_L$ is destroyed by $\delta_g\neq 0$.
An interesting property of $\xi_m$ is its sensitivity to two types of asymmetry that combine in a non interfering way.
The correction of $\delta_g\neq 0$ can be implemented in $x_{g4}$ at the beginning of the calculation of $\xi_m$ or implemented at the end (subtracting its value from $\xi_m$ ). These two
different procedures give identical results. This property resembles a linear combination of effects.

Similar analysis performed on the normal strip sensors gives analog results.
The quality of the $\theta_L$ determination has a similar precision, its resolution
is better than that of $\delta_g$.

\section{Conclusions}
The properties of the COG algorithms are able to access at a very detailed aspects of the detector:
the COG position of the response function and the Lorentz angle. The direct measurement is sufficiently
complex and could be unnecessary in many typical case.
This extraction can estimate these parameters from the data acquired in standard test
beam (or in a running experiment), with a simpler requirement of precise angular positioning of
the detector.
The noise introduces perturbation, but a strong relation to the
asymmetry is saved even in the worst case. The Lorentz angle determination shows a
modest sensitivity to the noise.

The present procedure is able to separate the intrinsic asymmetry of the $\varphi(\varepsilon)$
and the induced asymmetry due to incorrect initial conditions. The minimal asymmetry should be
the intrinsic one, but it is conceptually difficult to separate the two.
In spite of this,
the simulations show an excellent ability
to detect $\delta_g$ in the noiseless cases, giving to the phase
shifts of equation~\ref{eq:equation28} a robust meaning in the explored
range of asymmetry.  The noise modifies this picture adding a blurring in the reconstructions
that perturbs the efficacy of equation~\ref{eq:equation28}. But, the moderate noise of the
floating strip side has a negligible effect on $\delta_g$. In the simulations,
the $x_{g4}$ has a RMS of $4.2 \mu m$ on a strip pitch of $51 \mu m$.
For the normal strips, the noise is drastically higher (a RMS
of $10.6 \mu m$ on a strip pitch of $63 \mu m$) and $\delta_g$ estimation
appreciably degrades. The noise tends to mask the effect of the asymmetry
adding deformations that round  $\varphi(x)$ with a decreasing
of the resulting $\delta_g$.

The indicator of a non zero $\delta_g$ is the average of $x_{gk}$ at
orthogonal incidence. In the two cases we explored, this average
has a quite different relation to the asymmetry. In the first
case, to a large $x_{gk}$-averages corresponds a small asymmetry,
the reverse in the second case. Equation~\ref{eq:equation10}, as
in the example of figure~\ref{fig:figura03_a}, allows a visual
inspection of $\varphi(x)$. For normal strips, the noise in the
$x_{g4}$-algorithm masks almost completely the asymmetry, but the two strip
algorithm is able to give interesting results.

The asymmetries we consider have their principal effect on the
central strip. The capacitive coupling introduces long range
interactions in the nearby strips, and these interactions can
be asymmetric. The $x_{gk}$-averages are sensible to very small
effects and they may signal even long range asymmetries.
Equation~\ref{eq:equation10} is not fit to handle these
effects, it overlaps the tails of the
$\varphi(x)$ outside a strip range creating fake distortions.
Assembly of strips must be explored if
an indication of these long range effects is acquired.

The robustness of the approach is tested at non orthogonal incidence angle. The
parameter $\xi_m$ shows a surprising strict linear behavior, in the case of floating
strip sensor, that allows an increase of precision with a linear interpolation
of the data. For the normal strip case, appreciable deviation from linearity are
observed, but, even in this case an interpolation with a low degree polynomial
has beneficial effects on the $\delta_g$ determination.

The simulations at non orthogonal incidence suggest that the approach can be used for the
Lorentz angle determination. The approximation of the magnetic field effect as an
effective rotation of the reference system is probably very rough, in any case
$\xi_m$ is able to detect the angle of maximal symmetry with an excellent precision.
In the case of $\delta_g=0$ other simpler indicators has a comparable sensitivity
to the maximal symmetry: the averages of $x_{g4}$ and of $x_{g5}$ cross the $\theta=0$
line at the maximal symmetry.
These indicators become useless in presence of small $\delta_g\neq 0$. On
the contrary $\xi_m$  saves its efficiency to detect the maximal symmetry
even in presence of $\delta_g\neq 0$. With a first set of measurements
without the magnetic
field, $\delta_g$ can be measured and this correction must be implemented
in the $\xi_m$ calculation on the data with the magnetic field/effective rotation. Now
$\xi_m$ crosses the $\theta=0$ line at $-\theta_L$ as expected (with
our angle definitions). The two effects of the effective rotation and
$\delta_g\neq 0$ look to combine in an almost independent way. In fact,
the correction $\delta_g$ can be used to correct $x_{g4}$
or $x_{g5}$ before calculating $\xi_m$, or the correction $\delta_g$ can
be applied
directly at the end of the $\xi_m$ determination with identical results.

All these
simulations assume small values of $\delta_g$ and $\theta_L$. It is evident that
the explored values of $\delta_g$ are larger than these we can expect from the
detectors; the limitation to $\theta_L$ are easily overcome working around the
expected $\theta_L$ to have an its precision determination where the standard
methods have a low sensibility.



\end{document}